# The coexistence and evolution of attractors in the web map with weak dissipation

*A.V. Savin, D.V. Savin*

*Saratov State University, Saratov, Russia*

*E-mail: AVSavin@rambler.ru*

*The dynamics of the web map with weak linear dissipation is studied. The evolution of the co-existing attractors and the structure of their basins while changing the dissipation and nonlinearity are revealed. It is shown that the structure of the basins remains the same when the dissipation and nonlinearity changes simultaneously.*

Nowadays there are a number of works (e.g. [1–8]) concerning the dynamics of systems with weak dissipation. Such systems demonstrate intermediate behavior revealing features similar to conservative and dissipative dynamics. The main feature of it is a coexistence of a large number of low-period periodic attractors.

The main part of existing papers consider the effect of weak dissipation on the phase space of typical Hamiltonian systems which coincide the condition of the KAM theorem. But it is well known that degenerate non-integrable Hamiltonian systems demonstrate the special type of structures in the phase space, called the stochastic web by G.M. Zaslavsky [9]. It should be mentioned that such systems can describe different physical processes associated with the motion of charged particles if electric and magnetic fields. Particularly, it was shown [10] that it can describe the dynamics in the semiconductors super-lattices. So it seems to be interesting to investigate the effect of weak dissipation on the phase space of degenerate systems because it differs significantly from those of non-degenerate systems.

### 1. The stochastic web in conservative systems

The simplest example of degenerate Hamiltonian system demonstrating stochastic web in the phase space is a pulse driven harmonic oscillator:

$$\ddot{x} + \omega_0^2 x = -\frac{\omega_0 K}{T} \sin x \sum_{n=-\infty}^{+\infty} \delta(t - nT) \quad (1)$$

It was investigated by G.M. Zaslavsky [ ] who obtained the precise stroboscopic map, usually called the web map:

$$\begin{aligned} x_{n+1} &= x_n \cos\frac{2\pi}{q} + (y_n + K \sin x_n)\sin\frac{2\pi}{q}, \\ y_{n+1} &= (y_n + K \sin x_n)\cos\frac{2\pi}{q} - x_n \sin\frac{2\pi}{q}, \end{aligned} \quad (2)$$

where $x_n$ and $y_n$ means the coordinates and velocity of the oscillator before the *n*-th pulse and the parameter $q=\omega_0 T/2\pi$ determines the order of the resonance. The stochastic web exists for natural *q* (except of 1 and 2), see Fig.1 for example. There are two different situations called the cases of crystal and quasi-crystal symmetry. The first takes place when *q* is 3, 4 or 6 and all hyperbolic fixed points are situated at one energy level so they have common separatrix, and the stochastic web exists at infinitely small nonlinearity due to the destruction of this separatrix. In the case of quasi-crystal symmetry the energy levels for different hyperbolic fixed points differs slightly so there is no common separatrix and the stochastic web occurs at very small but finite value of the nonlinearity.

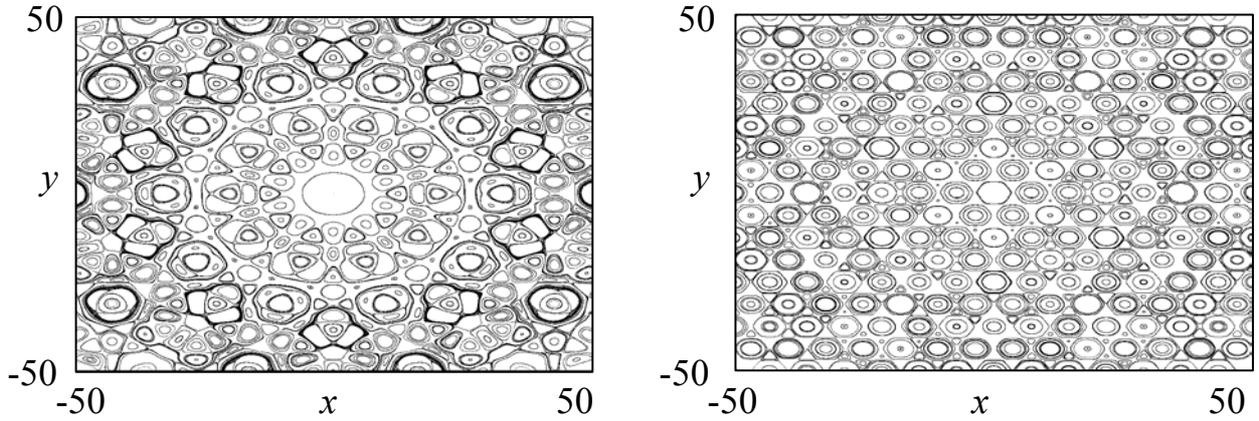

Fig.1. The trajectories for the number of initial points for the map (2) demonstrating the stochastic web with quasi-crystal (left, *q*=5) and crystal (right, *q*=6) symmetry. Nonlinearity parameter *K*=0.1.

## 2. The dissipative web map

Now let us introduce the weak linear dissipation in this system and consider the dissipative linear pulse driven oscillator:

$$\ddot{x} - 2\gamma\dot{x} + \omega_0^2 x = -\frac{\omega_0 K}{T}\sin x \sum_{n=-\infty}^{+\infty}\delta(t-nT) \qquad (3)$$

Due to the existence of the precise solution of the dissipative linear oscillator we obtain the precise stroboscopic map which is as follows:

$$\begin{aligned}
x_{n+1} &= e^{-\gamma\frac{2\pi}{q}}(x_n \cos\frac{2\pi}{q} + (y_n + K\sin x_n + \gamma x_n)\sin\frac{2\pi}{q}), \\
y_{n+1} &= e^{-\gamma\frac{2\pi}{q}}((y_n + K\sin x_n)\cos\frac{2\pi}{q} - (x_n(1+\gamma^2) + \gamma(y_n + K\sin x_n))\sin\frac{2\pi}{q}).
\end{aligned} \qquad (4)$$

We'll refer it as the *dissipative web map*. Here $\gamma$ is the dissipation parameter and the amplitude of the signal *K* in fact is the nonlinearity parameter, because the systems becomes linear at *K* equal to zero.

Fig. 2 demonstrates the attractors of map (4) plotted over the phase portrait of conservative map (2) with the same values of the parameters $K$ and $q$. We can divide all attractors into the two classes: "main" attractor, which is situated in the centre of plane, and the secondary attractors which are situated in the centers of the web cells. The secondary attractors disappear due to saddle-node bifurcation when the dissipation increases.

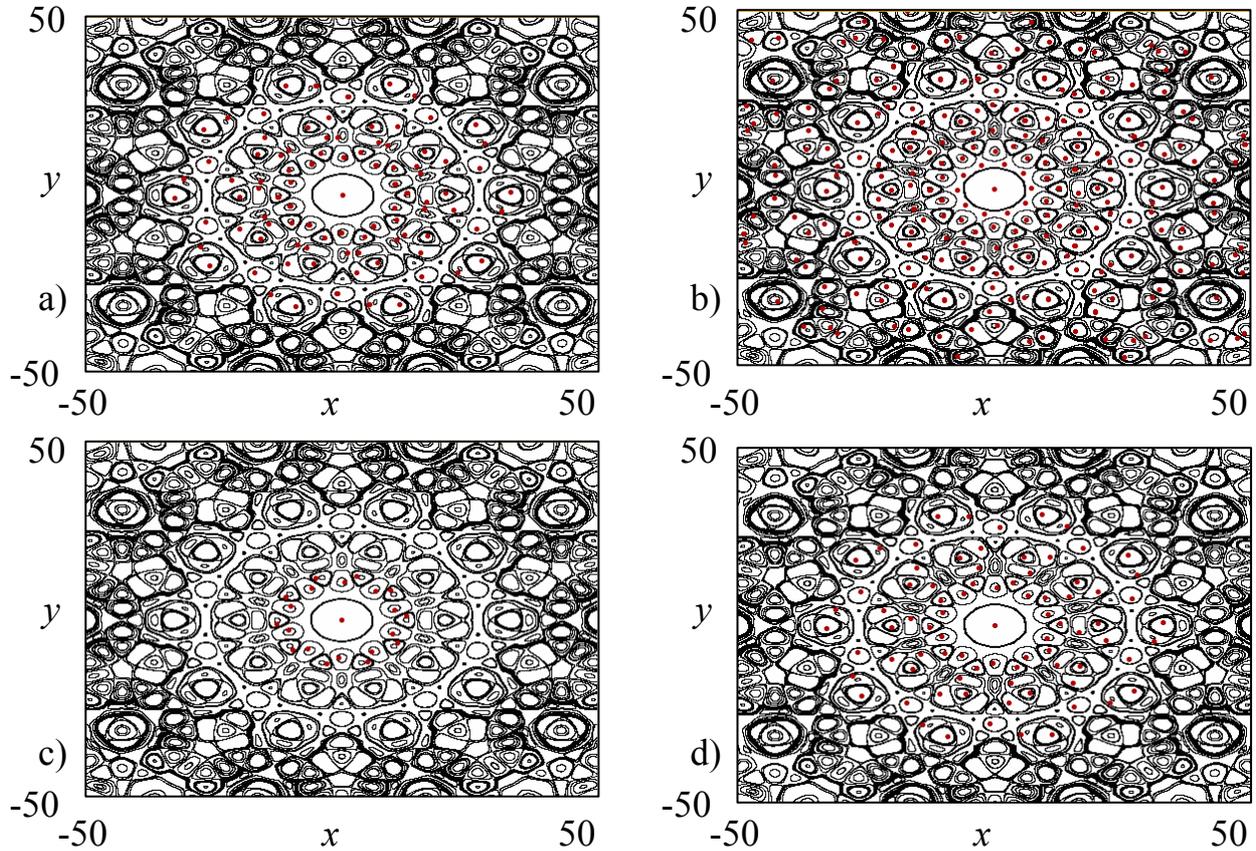

Fig.2. The attractors of the map (4) (marked with red points) plotted over the stochastic web for the map (2) with the same values of all parameter except of $\gamma$. The values of parameters are as follows: $q=5$; $K=0.05$, $\gamma=0.0005$ (a); $K=0.1$, $\gamma=0.0005$ (b); $K=0.05$, $\gamma=0.001$ (c); $K=0.1$, $\gamma=0.001$ (d).

The number of coexisting attractors decreases with the growth of dissipation and increases with the growth of the nonlinearity. The first fact seems to be rather trivial but the second is rather interesting. Moreover, one can see that the number and the disposition of attractors is practically the same if the value $K/\gamma$ remains the same (compare Fig. 2a and 2d).

Fig. 3 shows the basins of "main" and "secondary" attractors for different values of the nonlinearity parameter. The basins of the main attractors consist of the central cell of the web and several radial rays. These rays are divided into thinner rays as the nonlinearity increases and becomes very thin and riddled for sufficiently large nonlinearity (Fig. 3d) so the basin of main attractor reduces to the central cell of the web.

Now let us change the dissipation and nonlinearity simultaneously, so that $K/\gamma$ remains constant. In this case the structure of the basins remains exactly the same in the wide range of parameters (see Fig. 4, also compare it with Fig. 3a). Only at nonlinearity $K$ values near 1 the basin of the main attractor becomes riddled, and for the large values of $K$ (Fig. 4d) the basins of main and secondary attractors are completely riddled and the basin of the main attractor dominates.

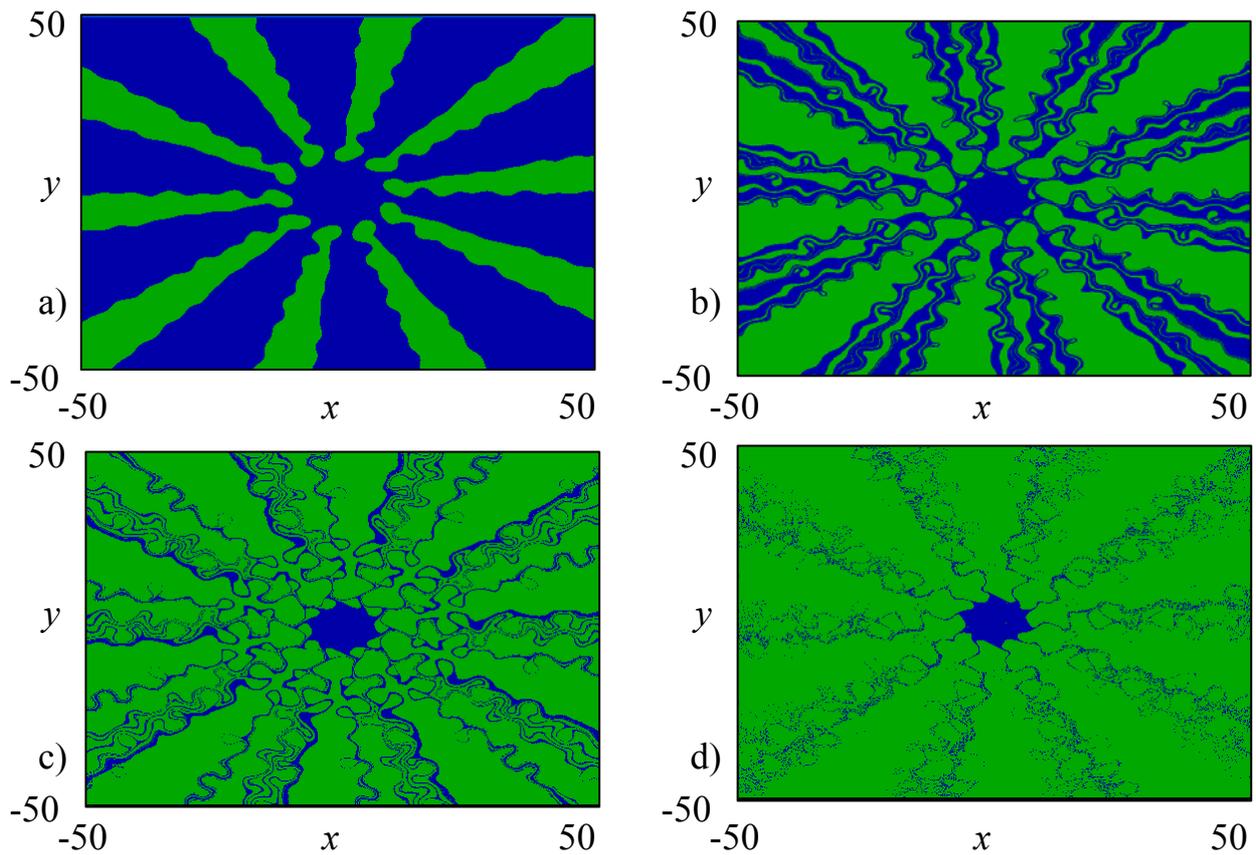

Fig.3. The basins of "main" (blue) and "secondary" (green) attractors of the map (4) for $q=5$, $\gamma=0.002$ and different values of the nonlinearity parameter: $K=0.1$ (a); $K=0.25$ (b); $K=0.5$ (c); $K=0.75$ (d).

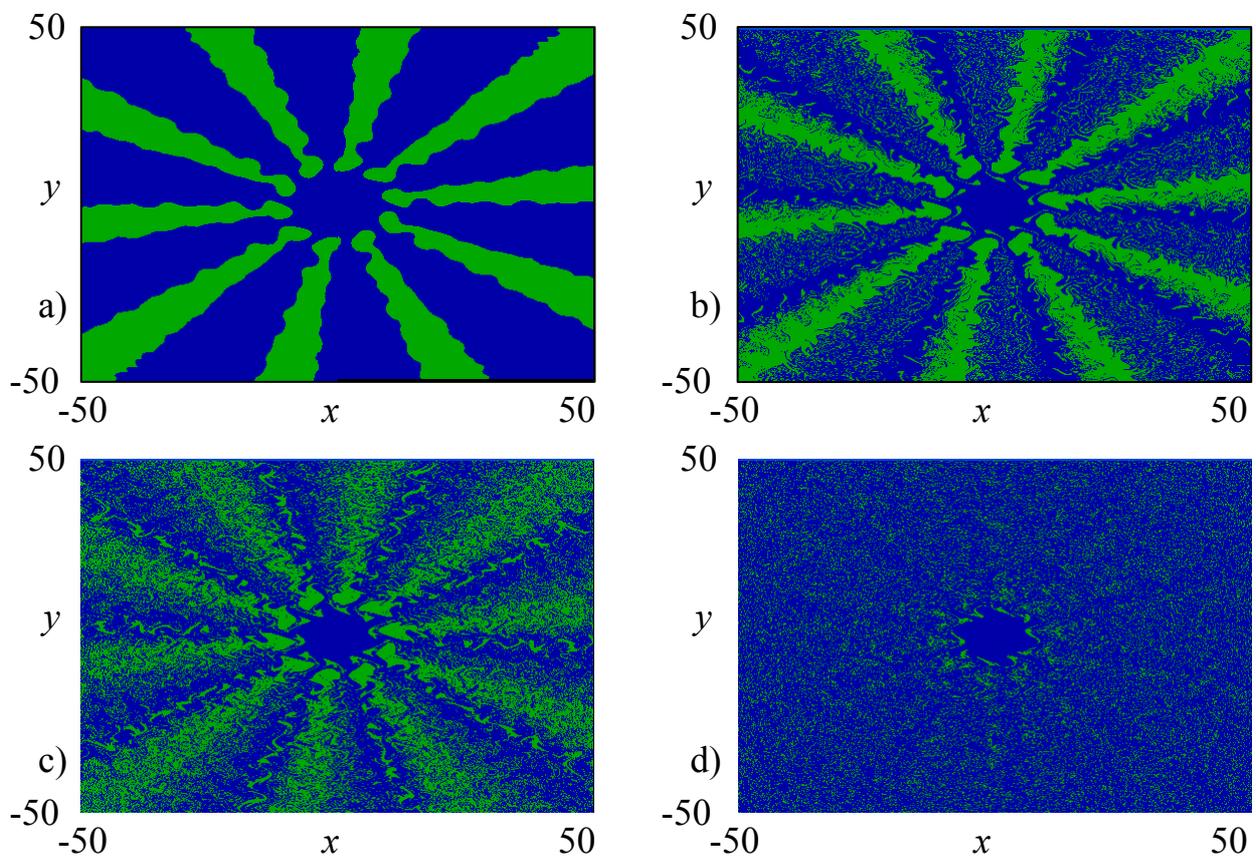

Fig.4. The basins of "main" (blue) and "secondary" (green) attractors of the map (4) for $q=5$, fixed ratio $K/\gamma=50$ and different values of the nonlinearity parameter: $K=0.4$ (a); $K=0.8$ (b); $K=1.1$ (c); $K=1.45$ (d).

It should be mentioned that nearly the same peculiarities can be observed for the webs with a crystal symmetry (i.e., for $q=3$, see Fig. 5) but the basin of the secondary attractor is larger in this case.

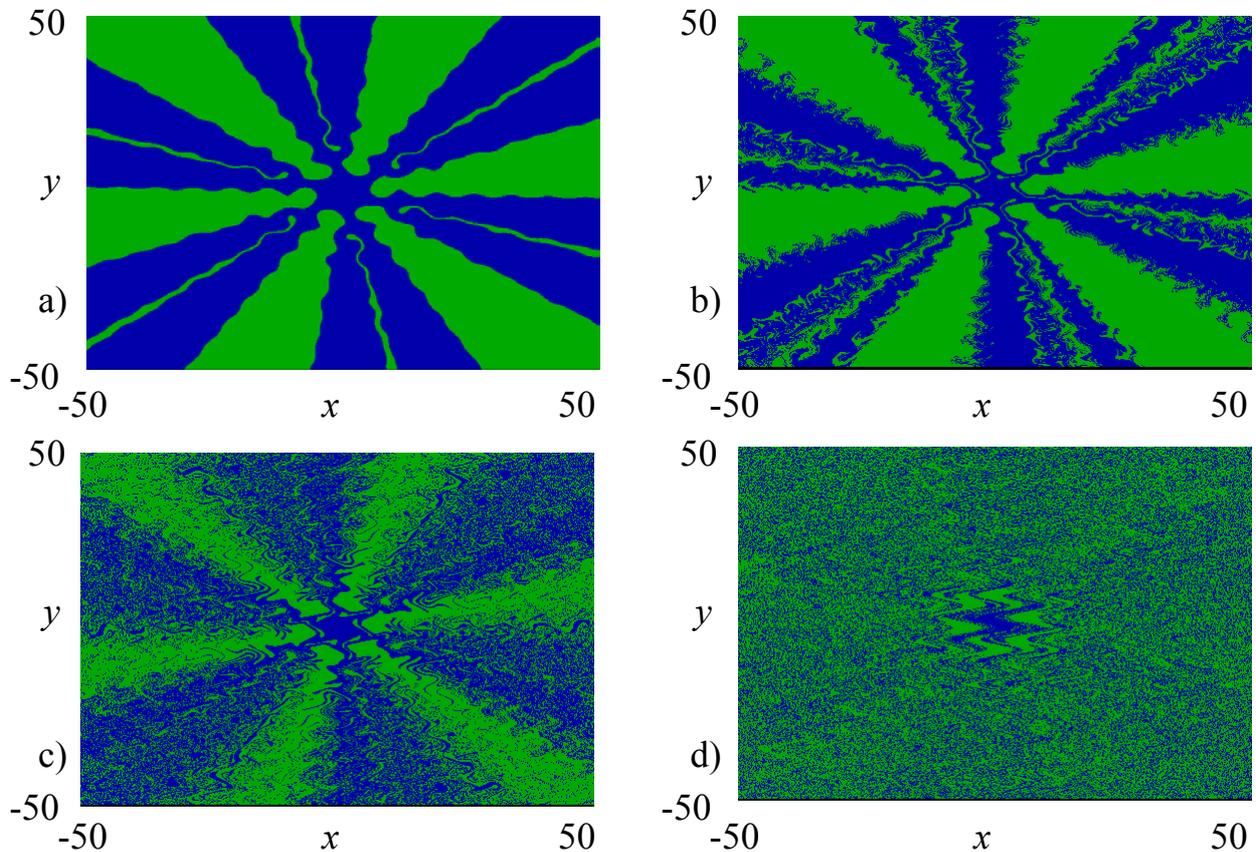

Fig.5. The basins of "main" (blue) and "secondary" (green) attractors of the map (4) for $q=3$, fixed ratio $K/\gamma=200$ and different values of the nonlinearity parameter: $K=0.2$ (a); $K=0.9$ (b); $K=1.3$ (c); $K=3.45$ (d).

We can explain this rather unusual feature as follows. The dynamics in the phase space is governed by two competitive processes: the diffusion along the stochastic web and the condensation to the attractor. The width of the stochastic web growth with the increase of the nonlinearity approximately as $exp(-const/K)$ [9] so the velocity of the diffusion increases. On the other hand, the time of diffusion becomes shorter approximately as $exp(-\gamma)$ when the dissipation increases because the transition to the attractor takes less time. So we can assume that when dissipation and nonlinearity increase simultaneously the effective shift of the point in the phase space which is the production of the velocity over time remains the same. The riddling of the basins at large nonlinearity values can be explained by the destruction of the web tori in the cells of the web – the well known phenomenon for the conservative systems.

**Conslusion**

So we have shown that the structure of the phase space of the dissipative web map including the location of attractors and the structure of their basins is governed by the ratio of nonlinearity to dissipation parameters for both crystal and quasi-crystal cases.

**Acknowledgments**

The work was supported by the Russian Foundation for Basic Researches, grant 12-02-31089.